\documentclass[prl,aps,twocolumn,superscriptaddress]{revtex4}
\usepackage{amsmath,amssymb}
\usepackage{graphicx}

\begin{document}

\title{Spontaneous Coherence in a Cold Exciton Gas}
\author{A.A.~High}
\affiliation{Department of Physics, University of California at San Diego, La Jolla, CA 92093-0319, USA}
\author{J.R.~Leonard}
\affiliation{Department of Physics, University of California at San Diego, La Jolla, CA 92093-0319, USA}
\author{A.T.~Hammack}
\affiliation{Department of Physics, University of California at San Diego, La Jolla, CA 92093-0319, USA}
\author{M.M.~Fogler}
\affiliation{Department of Physics, University of California at San Diego, La Jolla, CA 92093-0319, USA}
\author{L.V.~Butov}
\affiliation{Department of Physics, University of California at San Diego, La Jolla, CA 92093-0319, USA}
\author{A.V.~Kavokin}
\affiliation{School of Physics and Astronomy, University of Southampton, SO17 1BJ, Southampton, United Kingdom}
\author{K.L.~Campman}
\affiliation{Materials Department, University of California at Santa Barbara, Santa Barbara, CA 93106-5050, USA}
\author{A.C.~Gossard}
\affiliation{Materials Department, University of California at Santa Barbara, Santa Barbara, CA 93106-5050, USA}
\date{\today}

\begin{abstract}
\noindent Excitons, bound pairs of electrons and holes, form a model system to explore the quantum physics of cold bosons in solids. Cold exciton gases can be realized in a system of indirect excitons, which can cool down below the temperature of quantum degeneracy due to their long lifetimes. Here, we report on the measurement of spontaneous coherence in a gas of indirect excitons. We found that extended spontaneous coherence of excitons emerges in the region of the macroscopically ordered exciton state and in the region of vortices of linear polarization. The coherence length in these regions is much larger than in a classical gas, indicating a coherent state with a much narrower than classical exciton distribution in momentum space, characteristic of a condensate. We also observed phase singularities in the coherent exciton gas. Extended spontaneous coherence and phase singularities emerge when the exciton gas is cooled below a few Kelvin.
\end{abstract}

\maketitle

\textbf{1. Spontaneous coherence of excitons}

If bosonic particles are cooled down below the temperature of quantum degeneracy they can spontaneously form a coherent state in which individual matter waves synchronize and combine. Spontaneous coherence
of matter waves forms the basis for a number of fundamental phenomena in physics, including superconductivity, superfluidity, and Bose-Einstein condensation (BEC) \cite{Cornell02, Ketterle02, Abrikosov04, Ginzburg04, Leggett04}. Spontaneous coherence is the key characteristic of condensation in momentum space \cite{Penrose56}.

Excitons are hydrogen-like bosons at low densities \cite{Keldysh68} and Cooper-pair-like bosons at high densities \cite{Keldysh65}. The bosonic nature of excitons allows for condensation in momentum space, i.e. emergence of spontaneous coherence. Designing semiconductor structures with required characteristics and controlling the parameters of the exciton system gives an opportunity to study various types of exciton condensates.

A condensate of exciton-polaritons was recently realized in semiconductor microcavities \cite{Deng02, Kasprzak06, Christopoulos07, Bajoni07, Love08}. This condensate is characterized by a strong coupling of excitons to the optical field and a short lifetime of the polaritons. Unlike BEC, equilibrium is not required for the polariton condensation and coherence in the polariton condensate forms due to a coherent optical field similar to coherence in lasers \cite{Littlewood04}.

There are intriguing theoretical predictions for a range of coherent states in cold exciton systems, including BEC \cite{Keldysh68}, BCS-like condensation \cite{Keldysh65}, charge-density-wave formation \cite{Chen91}, and condensation with spontaneous time-reversal symmetry breaking \cite{Wu08}. In these condensates, spontaneous coherence of exciton matter waves emerges below the temperature of quantum degeneracy.

Since excitons are much lighter than atoms, quantum degeneracy can be achieved in excitonic systems at temperatures orders of magnitude higher than the micro-Kelvin temperatures needed in atomic vapors \cite{Cornell02, Ketterle02}. Exciton gases need be cooled down to a few Kelvin to enter the quantum regime. Although the temperature of the semiconductor crystal lattice $T_{\text{lat}}$ can be lowered well below $1\,\text{K}$ in He-refrigerators, lowering the temperature of the exciton gas $T_X$ to even a few Kelvin is challenging \cite{Tikhodeev98, Jang06, Ideguchi08}. Due to recombination, excitons have a finite lifetime which is too short to allow cooling to low temperatures in regular semiconductors. In order to create a cold exciton gas with $T_X$ close to $T_{\text{lat}}$, the exciton lifetime should considerably exceed the exciton cooling time. Besides this, the realization of a cold and dense exciton gas requires an excitonic state to be the ground state and have lower energy than the electron-hole liquid \cite{Keldysh86}.

\begin{figure*}[tbp]
\centering
\includegraphics[width=15cm]{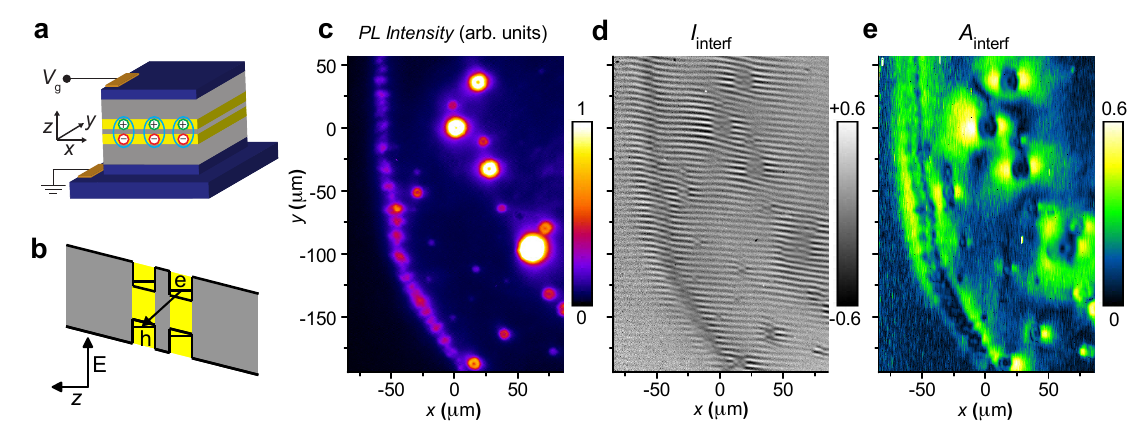}
\caption{Emission, interference, and coherence patterns of indirect excitons. (a) Schematic of CQW structure: $n^+$-GaAs (blue), Al$_{0.33}$Ga$_{0.67}$As (gray), GaAs QW (yellow). (b) CQW band diagram. (c) Emission pattern. (d) Interference pattern for $\delta x = -2\, \mu\text{m}$. (e) Pattern of amplitude of the interference fringes presenting a map of coherence. $T_{\text{bath}} = 0.1\,\text{K}$.}
\end{figure*}

A gas of indirect excitons fulfills these requirements. An indirect exciton can be formed by an electron and a hole confined in separate quantum well layers (Fig. 1a,b). The spatial separation allows one to control the overlap of electron and hole wavefunctions. This way, one can engineer indirect excitons with radiative lifetimes and spin relaxation times orders of magnitude longer than those of regular excitons \cite{Lozovik76, Fukuzawa90, Maialle93}. Due to their long lifetimes, indirect excitons can cool down well below the temperature of quantum degeneracy \cite{Butov01}.

In earlier studies, evidence for spontaneous coherence was obtained for indirect excitons in coupled quantum wells (CQW) \cite{Butov98} and for indirect excitons in quantum Hall bilayers \cite{Spielman00, Eisenstein04}. The onset of spontaneous coherence was evidenced by a strong enhancement of the recombination \cite{Butov98} and tunneling \cite{Spielman00} rate, respectively. The results of other transport and optical experiments were also consistent with spontaneous coherence of indirect excitons \cite{Butov98, Eisenstein04, Butov94, Butov01, Tutuc04, Tiemann08, Croxall08, Seamons09, Karmakar09}. However, no direct measurement of coherence has been performed in these studies.

Exciton coherence is imprinted on coherence of their light emission, which one can study by interferometry. In our earlier work, we reported an enhancement of the exciton coherence length in the macroscopically ordered exciton state (MOES) \cite{Yang06, Fogler08}. However, these experiments used a single-pinhole interferometric technique which does not measure the coherence function $g_1(\delta \mathbf{r)}$ and the derivation of the exciton coherence length in \cite{Yang06, Fogler08} was based on a mathematical analysis of the data.

Here, we report on the direct measurement of spontaneous coherence in a gas of indirect excitons in CQW. The studied indirect excitons may have four spin projections on the $z$ direction normal to the CQW plane: $J_z=-2,-1,+1,+2$. The states $J_z=-1$ and $+1$ contribute to left- and right-circularly polarized emission and their coherent superposition -- to linear polarized emission, the states $J_z=-2$ and $+2$ are dark \cite{Maialle93,Wu08}. The exciton condensate is a four-component coherent state in general. The build-up of exciton coherence should manifest itself in an increase of the coherence length and polarization degree of the exciton emission. The former phenomenon is general for both one- and multi-component condensates \cite{Penrose56} while the latter is specific to multi-component condensates \cite{Read09}. In this work, we report the emergence of both long-range spontaneous coherence of excitons and spontaneous polarization. A pattern of extended spontaneous coherence, measured by shift-interferometry, is correlated with a pattern of spontaneous polarization, measured by polarization-resolved imaging. These two experiments reveal the properties of a multi-component coherent state.

\vskip 5mm \textbf{2. Experiment}

\begin{figure*}[tbp]
\centering
\includegraphics[width=16cm]{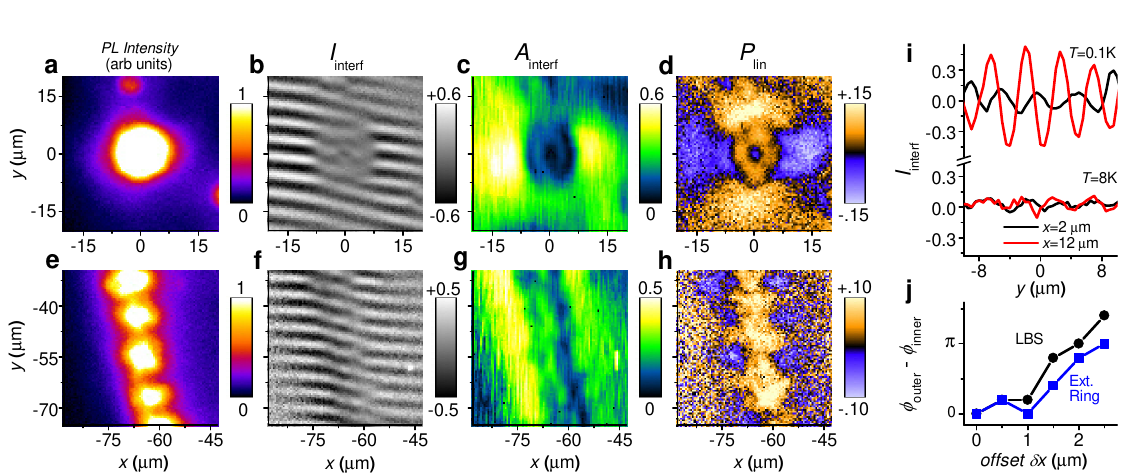}
\caption{Coherence of indirect excitons in regions of an LBS and the external ring. (a-d) refer to the region of an LBS. (e-h) refer to a region of the external ring. (a,e) Emission pattern. (b,f) Interference pattern at shift $\protect\delta x = -2 \,\mu\text{m}$. (c,g) Amplitude $A_{\text{interf}}(x,y)$ of interference fringes. (d,h) Linear polarization of exciton emission $P_{\text{lin}} = (I_x - I_y)/(I_x + I_y)$. (i) $y$ axis cross sections of $I_{\text{interf}}(x,y)$ at $x = 2 \,\mu\text{m}$ (black lines) and $x = 12 \,\mu\text{m}$ (red lines) at $T_{\text{bath}} = 0.1\,\text{K}$ and $8\,\text{K}$. (j) The shift in the phase of interference fringes in (b) at $r \approx 7\, \mu\text{m}$ (black) and in (f) at $\approx 4\, \mu\text{m}$ away from the middle circumference of the external ring (blue) \textit{vs.} $\protect\delta x$. $T_{\text{bath}} = 0.1\,\text{K}$ for (a-h,j).}
\end{figure*}

The pattern of the first-order coherence function $g_1(\delta x)$ is measured by shift-interferometry: the emission images produced by arm 1 and 2 of the Mach-Zehnder interferometer are shifted with respect to each other to measure the interference between the emission of excitons separated by $\delta x$.

A cold gas of indirect excitons is realized in a GaAs/AlGaAs CQW structure (Figs.~1a,b). Long lifetimes of the indirect excitons allow them to cool to low temperatures within about $0.1\,\text{K}$ of the lattice temperature, which can be lowered to $0.1\,\text{K}$ in an optical dilution refrigerator. This allows the realization of a cold exciton gas with temperature well below the temperature of quantum degeneracy, which is in the range of a few Kelvin for typical exciton densities $\sim 10^{10}\,\text{cm}^{-2}$ for the studied CQW \cite{Butov01}.

In the reported experiments, extended spontaneous coherence is observed in the region of rings in the exciton emission pattern. Exciton rings, including the inner ring, external ring, and localized bright spot (LBS) rings were observed earlier \cite{Butov02}. The external and LBS rings are exciton sources that form on the boundaries between electron-rich and hole-rich regions; the former is created by current through the structure (specifically, by the current filament at the LBS center in the case of the LBS ring), while the latter is created by optical excitation \cite{Butov04, Rapaport04, Haque06, Yang10}. Figure~1c shows a segment of the exciton emission pattern with a section of the external ring and smaller LBS rings. The rings provide optimal conditions for the formation of a cold exciton gas for the following reasons: (i) the rings are spatially separated from the optical excitation spot and, therefore, the excitons in the rings are formed from cold electrons and holes thermalized to the lattice temperature, (ii) the heating of the exciton gas due to the binding energy released at the exciton formation is suppressed due to long lifetimes of indirect excitons, (iii) indirect excitons generated in the rings can further cool down to the lattice temperature when they travel out of the source. The studies of exciton kinetics \cite{Hammack09} show that indirect excitons cool down to the lattice temperature within first few microns from the source and the cold indirect excitons can travel over tens of microns out of the source before recombination.

In this work, the area of study is beyond the inner ring, which is formed by indirect excitons traveling away from the optical excitation spot \cite{Butov02, Ivanov06, Hammack09}. The photoexcited electrons recombine within the inner ring \cite{Hammack09, Ivanov06}. Therefore, in the area of external and LBS rings with no photoexcited electrons coherence forms spontaneously; it is not affected by coherence of the optical pumping.

\vskip 5 mm \textbf{3. Pattern of spontaneous coherence}

Figure~1 presents the pattern of interference fringes (Fig.~1d) and the map of their amplitude $A_{\text{interf}}$ (Fig.~1e). As detailed in Sec.~4, $A_{\text{interf}}$ describes the coherence degree of excitons. The regions of extended spontaneous coherence of excitons correspond to green color in this figure. The observed properties of spontaneous coherence of excitons are discussed below.

Figure~2 presents the patterns of coherence of indirect excitons in regions of an LBS and external ring. The observed properties of exciton coherence are qualitatively similar around both these sources of cold excitons. We first consider an LBS region. At low temperatures, a strong enhancement of $A_{\text{interf}}$ is observed at distance $r \sim r_0 = 7\, \mu\text{m}$ away from the LBS center (Figs.~2b,c,i). Furthermore, the phase of the interference fringes $\phi_{\text{interf}}$ experiences a shift at $r \sim r_0 = 7\, \mu\text{m}$, which defines a phase domain boundary (Fig.~2b). The shift in phase correlates with the enhancement of $A_{\text{interf}}$ (Fig.~2b,c). Its magnitude $\delta \phi = \phi_{\text{interf}}^{\text{outer}} - \phi_{\text{interf}}^{\text{inner}}$ increases with $\delta x$ (Fig.~2j). Since for excitons propagating with momentum $q$ its phase evolves in space as $q r$, the shift in the phase of the interference fringes implies a jump in (average) exciton momentum $\delta q \sim \delta \phi / \delta x \sim 2 \,\mu\text{m}^{-1}$ at $r = r_0$.

Figure~2d presents a pattern of linear polarization around an LBS. It spatially correlates with the pattern of the amplitude and phase of the interference fringes, cf.~Figs.~2b, c, and d. At $r \gtrsim r_0$ a vortex of linear polarization with the polarization perpendicular to the radial direction is observed. Such polarization vortices appear due to precession of the Stokes vector for excitons propagating out of the LBS origin \cite{High11}. The polarization vortices emerge at low temperatures along with spontaneous coherence of excitons.

Similar phenomena are observed in the external ring region. At low $T$, the MOES forms in the external ring \cite{Butov02} (Figs.~1c and 2e) and a periodic polarization texture forms around the periodic array of beads in the MOES \cite{High11} (Fig.~2h). Figures 2f and 2g show the extended spontaneous coherence of excitons observed in the MOES. It emerges at low temperatures along with the spatial order of exciton beads and periodic polarization texture.

\vskip 5 mm \textbf{4. First-order coherence function}

Coherence of the exciton gas is directly characterised by the coherence of exciton emission, described by the first-order coherence function $g_1(\mathbf{\delta r})$. In turn, this function is given by the amplitude of the interference fringes $A_{\text{interf}}(\mathbf{\delta r})$ in `the ideal experiment' with perfect spatial resolution. In practice, the measured $A_{\text{interf}}(\mathbf{\delta r})$ is given by the convolution of $g_1(\mathbf{\delta r})$ with the point-spread function (PSF) of the optical system used in the experiment \cite{Fogler08}. The PSF width corresponds to the spatial resolution of the optical system.

The measurements of $A_{\text{interf}}(\delta x)$ in the polarization vortex and in the LBS center are presented in Fig.~3a. In the hot LBS center, $A_{\text{interf}}$ quickly drops with $\delta x$ and the shape $A_{\text{interf}}(\delta x)$ fits well to the PSF, which is presented by blue line. In the polarization vortex, $g_1(\delta x)$ extends to large $\delta x$ demonstrating extended spontaneous coherence. A fit to the experimental points computed using a model described below is shown by black line.

Figures~3b and 3c demonstrate the relation between the first-order coherence function and particle distribution in momentum space. Figure 3b presents $g_1(\delta x)$ for a classical gas (blue dashed line) and quantum gas (black dashed line) for a spatially homogeneous gas of noninteracting particles with a quadratic dispersion (see SI). Both gases are at $0.1\,\text{K}$, the classical gas has a very small occupation number $n_{0} \ll 1$ of the $q = 0$ momentum state while the quantum gas has $n_{0} = 5000$. The convolution of these $g_1(\delta x)$ with the PSF produces black and blue solid lines, which fit to $A_{\text{interf}}(\delta x)$ in the spin polarization vortex and in the LBS center, respectively (Figs.~3a,b). The Fourier transform of $g_1(\delta x)$ in Fig.~3b gives the momentum occupation factor $n_q$ in Fig.~3c.

\begin{figure}[tbp]
\centering
\includegraphics[width=8.5cm]{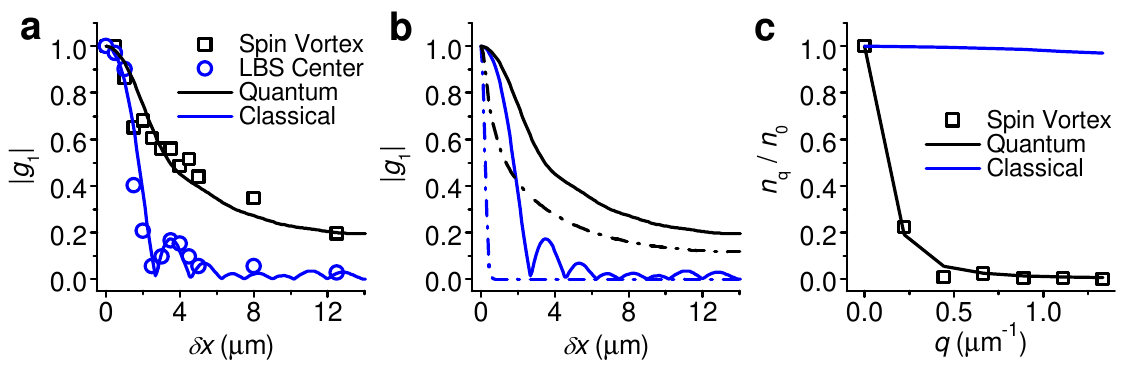}
\caption{First-order coherence function and distribution in momentum space. (a) Measured $|g_1(\delta x)|$ for the polarization vortex (squares) and LBS center (circles) and simulated $|g_1(\delta x)|$ for a quantum (black line) and classical (blue line) gas. (b) Simulated $|g_1(\delta x)|$ for a quantum (black) and classical (blue) gas with (solid) and without (dashed) convolution with PSF. (c) Distribution in momentum space obtained by the Fourier transform of $g_1$ in (b) for a quantum (black line) and a classical (blue line) gas.}
\end{figure}

Figures 3b and 3c illustrate that a classical gas is characterized by a broad distribution in momentum space $n_q$ and a narrow first-order coherence function $g_1(r)$, while a quantum gas is characterized by a narrow $n_q$ and an extended $g_1(r)$. For a classical gas, $g_1(r)$ drops within the thermal de~Broglie wavelength $\lambda_{dB}$, which scales $\propto T^{-1/2}$ and is about $0.5 \mu$m at $0.1\,\text{K}$. The extension of $g_1(r)$ well beyond $\lambda_{dB}$ indicates a coherent exciton state, characteristic of a condensate.

Figure 3a illustrates also why $\delta x = 2 \mu$m is selected for mapping extended spontaneous coherence of excitons. The shift $\delta x = 2 \mu$m is chosen to exceed both $\lambda_{dB}$ and the PSF width. At such $\delta x$, only weak coherence given by the PSF value at $\delta x = 2 \mu$m can be observed for a classical gas. The regions of enhanced coherence exceeding such background level reveal the regions with extended spontaneous coherence of excitons.

\vskip 5 mm \textbf{5. Pattern of coherence length}

\begin{figure}[tbp]
\centering
\includegraphics[width=8.5cm]{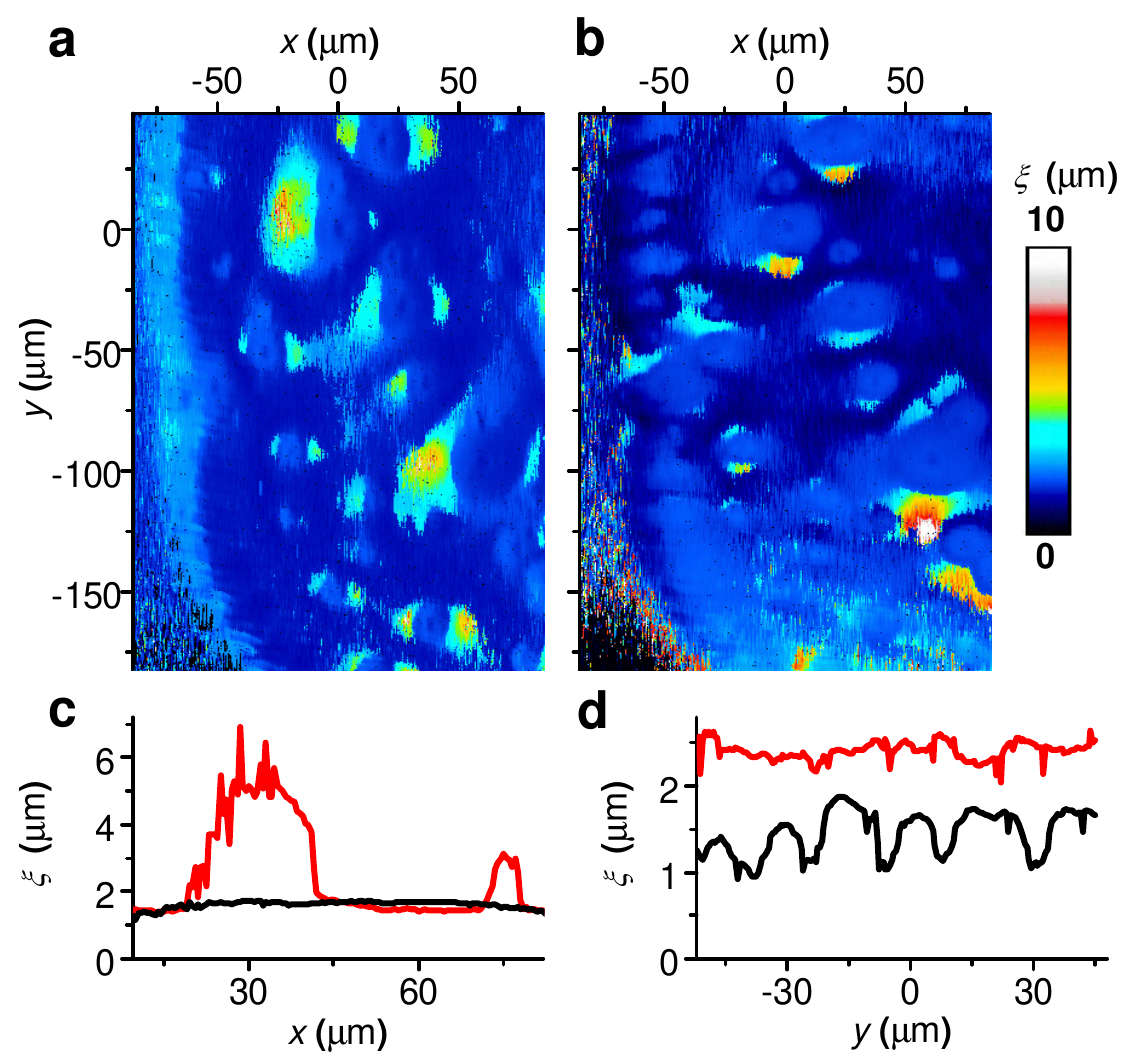}
\caption{Pattern of coherence length of excitons $\protect\xi(x,y)$. (a,b) $\xi(x,y)$ for a shift along (a) $x$ and (b) $y$. (c,d) Cross section of $\xi (x,y)$ (c) through LBS [$y = -90\, \mu\text{m}$] and (d) along external ring [$x = -70\, \mu\text{m}$] for a shift along $x$ (red) and $y$ (black). $T_{\text{bath}} = 0.1\,\text{K}$.}
\end{figure}

The spatial extension of $g_1(\delta r)$ can be characterized by a coherence length $\xi$. Here, to consider all points in the pattern on equal footing, we evaluate $\xi$ as $\delta r$ at which the interference visibility drops $e$ times. We measured the exciton interference pattern at different $\delta r$ to produce the spatial map of $g_1(\delta r)$ and, in turn, $\xi$. Figures~4a and 4b show the pattern of $\xi$ for the shift between the interfering excitons along $x$ and $y$, respectively. Figures~4c and 4d present the cross sections of $\xi(x,y)$ in the region of the polarization vortex (c) and MOES (d).

The regions of a classical gas in the $\xi(x,y)$ pattern correspond to the smallest observed coherence length given by the PSF width. Long-range spontaneous coherence of excitons is observed in the polarization vortices and in the macroscopically ordered exciton state (Fig.~4). The coherence length in these regions is much larger than in a classical gas, indicating a coherent state with a much narrower than classical exciton distribution in momentum space, characteristic of a condensate.

The observed coherence length in the polarization vortex exceeds $\lambda_{dB}=0.5\, \mu$m at 0.1 K by more than order of magnitude (Fig.~4). The coherence length in the MOES is smaller than in the polarization vortex. This may be related to fluctuations of the exciton density wave in the external ring. Such fluctuations were observed recently and their studies will be reported in future works.

The patterns of coherence length are different for the shifts along $x$ and $y$, revealing a directional property of exciton coherence. In the region of the polarization vortices, $\xi$ is higher in the direction along the shift between the interfering excitons [$x$ direction for the $\delta x$-shift (Fig.~4a), and $y$ direction for the $\delta y$-shift (Fig.~4b)]; In the region of the MOES, $\xi$ is higher for $\delta x$-shift along the direction of exciton propagation away from the external ring in Fig.~4a. These data indicate that the extension of $g_1(\mathbf{r)}$ is higher when the exciton propagation direction is along vector \textbf{r}.

\vskip 5 mm \textbf{6. Phase singularities: Fork-like defects in exciton interference pattern}

\begin{figure}[tbp]
\includegraphics[width=8.5cm]{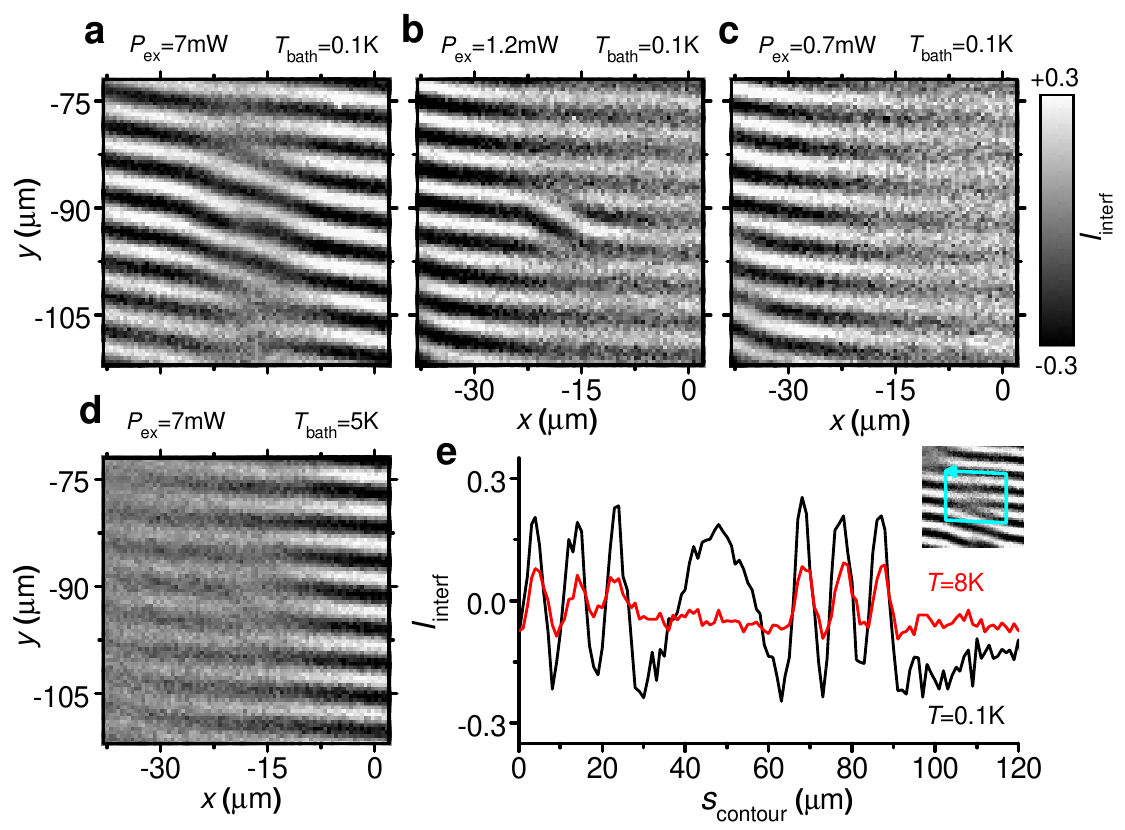}
\caption{Fork-like defects in exciton interference pattern. (a-d) Interference pattern $I_{\text{interf}}(x,y)$ around an LBS vs. (a-c) $P_{\text{ex}}$ and (a,d) $T_{\text{bath}}$. (e) $I_{\text{interf}}$ along a closed contour [shown in inset] for $T_{bath} = 0.1\,\text{K}$ (black) and $8\,\text{K}$ (red). $\protect\delta x = 2\,\mu\text{m}$.}
\end{figure}

A well known example of phase singularities is a quantized vortex. Studies of quantized vortices in various coherent states, from superconductors to atom BEC, have been an essential part of physics. In a singly quantized vortex, the phase of the wavefunction winds by $2\pi$ around the singularity point. Therefore, a fork-like defect in a phase pattern can be a signature of a quantized vortex. Fork-like defects in interference patterns were reported for optical vortices, see \cite{Scheuer99} and references therein, vortices in atom condensates \cite{Inouye01, Chevy01, Hadzibabic06}, and polariton vortices \cite{Lagoudakis08, Lagoudakis09}.

We observed a number of fork-like defects in the interference pattern of a cold exciton gas. For instance more than 20 of them are present in Fig.~1d. A larger scale is presented in Fig.~5. Figure 5a shows forks in the interference pattern indicating the presence of phase singularities. We study the properties of this phenomenon and show that its origin is different from a quantized vortex.

The forks in the interference pattern are observed at low temperatures in a quantum exciton gas (Fig.~5a) and vanish at high temperatures in a classical exciton gas (Fig.~5d). At low temperatures, a closed contour around the fork crosses an odd number of interference fringes so the phase of the interference fringes winds by $2\pi$, indicating a phase singularity (Fig.~5e). Similar properties are observed for quantized vortices.

However, the distance between the left- and right-facing forks in the interference pattern is not equal to the shift $\delta x = 2\, \mu\text{m}$ in the shift-interferometry experiment and can be controlled by the excitation power (Fig.~5a-c). This shows the difference between the observed phase singularity and a quantized vortex. Indeed, straightforward simulations show that a quantized vortex is characterized by a pair of opposite forks separated by the distance equal to the shift in the shift-interferometry experiment. This shows that forks in the interference pattern may correspond to objects different from quantized vortices.

Simulations of the interference pattern produced by a ring-shaped source, such as an LBS ring, result in the interference pattern with opposite forks separated by a distance much larger than $\delta x$, in qualitative agreement with the experiment (see SI). A ring-shaped source with particles propagating away from their origin on the ring produces a more complicated phase pattern than a vortex, yet both objects are characterized by the spreading of particle velocities over all directions. The observed phase singularities constitute the properties of a quantum exciton gas with extended spontaneous coherence (Fig.~5a) and no such phase singularity is observed at high temperatures in a classical gas (Fig.~5d).

\vskip 5 mm \textbf{7. Summary}

Spontaneous coherence is observed in a gas of indirect excitons. The coherence length in the macroscopically ordered exciton state and in the vortices of linear polarization reaches several microns, which is much larger than in a classical exciton gas, indicating a coherent state with a much narrower than classical exciton distribution in momentum space, characteristic of a condensate. A pattern of extended spontaneous coherence is correlated with a pattern of spontaneous polarization, revealing the properties of a multi-component coherent state. We also observed phase singularities including phase domains and fork-like defects in the interference pattern. All these phenomena emerge when the exciton gas is cooled below a few Kelvin.

We thank Leonid Levitov, Lu Sham, Ben Simons, and Congjun Wu for discussions. This work was supported by the DOE Office of Basic Energy Sciences under award DE-FG02-07ER46449. The development of spectroscopy in a dilution refrigerator were supported by ARO grant W911NF-08-1-0341 and NSF grant 0907349. MF is supported by the UCOP.

\setcounter{figure}{0}
\renewcommand{\thefigure}{S\arabic{figure}}
\setcounter{equation}{0}
\renewcommand{\theequation}{S\arabic{equation}}

\vskip 1cm \textbf{Supplementary Information}

\vskip 5mm \textbf{CQW structure and experimental setup}

Experiments were performed on a $n^+ - i - n^+$ GaAs/AlGaAs CQW structure grown by molecular-beam epitaxy. The $i$ region consists of a single pair of 8 nm GaAs QWs separated by a 4 nm Al$_{0.33}$Ga$_{0.67}$As barrier and surrounded by 200 nm Al$_{0.33}$Ga$_{0.67}$As barrier layers. The $n^+$ layers are Si-doped GaAs with $N_{Si} = 5 \times 10^{17}\,\text{cm}^{-3}$. The electric field in the $z$ direction is controlled by the external voltage applied between $n^+$ layers. The small in-plane disorder in the CQW is indicated by the emission linewidth $\approx 1\,\text{meV}$. The laser excitation is performed by a HeNe laser at $\lambda = 633\,\text{nm}$ with the excitation power $P_{\text{ex}} = 1.2\,\text{mW}$ for the data in Figs.~1-3 and $P_{\text{ex}} = 2.9\,\text{mW}$ for the data in Fig.~4 of the main text. The photoexcitation is more than $400\,\text{meV}$ above the energy of indirect excitons and the $10\, \mu\text{m}$-wide excitation spot is farther than $80\, \mu\text{m}$ away from both the LBS and the external ring. The $x$-polarization is along the sample cleavage direction within the experimental accuracy.

\begin{figure}[tbp]
\centering
\includegraphics[width=8.5cm]{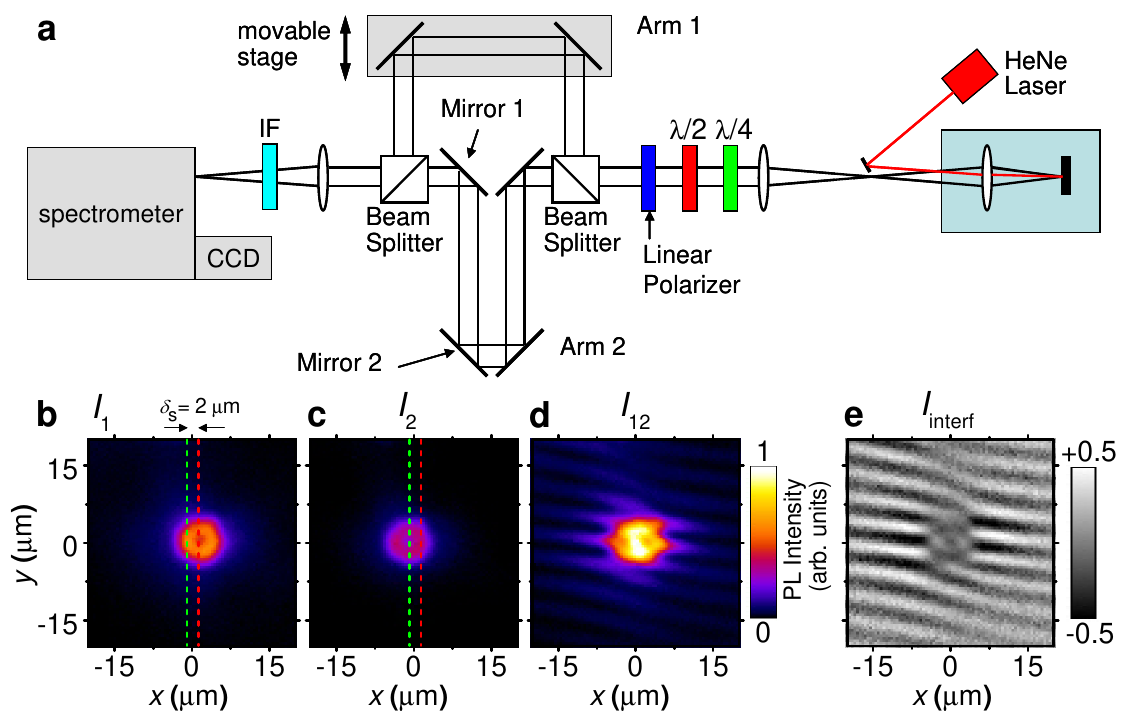}
\caption{Schematic of the experiment. (a) Interferometric setup. (b-d) Emission image of indirect excitons measured with (b) arm 1 open, (c) arm 2 open, and (d) both arms open. (e) $I_{\text{interf}}(x,y)$ obtained from (b-d). $P_{\text{ex}}=2.9\,\text{mW}$, $T_{\text{bath}} = 0.1\,\text{K}$.
\label{fig:schematic}
}
\end{figure}

We use a Mach-Zehnder (MZ) interferometer to probe the coherence of the exciton system (Fig.~\ref{fig:schematic}). The emission beam is made parallel by an objective inside the optical dilution refrigerator and lenses. A combination of a quarter-wave plate and a half-wave plate converts the measured polarization of the emission to the $y$-polarization, which is then selected by a linear polarizer. This ensures only the $y$-polarized light enters the MZ interferometer thus eliminating polarization-dependent effects, which otherwise can be caused by the interferometer and spectrometer. In the shift-interferometry experiment and polarization-resolved experiment, spectrometer operates in the zero-order dispersionless regime. The emission is split between arm 1 and arm 2 of the MZ interferometer. The path lengths of arm 1 and arm 2 are set equal using the movable stage. The interfering emission images produced by arm 1 and 2 of the MZ interferometer are shifted relative to each other along $x$ to measure the interference between the emission of excitons, which are laterally separated by $\delta x$. Mirror 1 and Mirror 2 are used to control the fringe period, and Mirror 2 is used to control $\delta x$. After the interferometer, the emission is filtered by an interference filter of linewidth $\pm 5 \text{nm}$ adjusted to the exciton emission wavelength $\lambda = 800\,\text{nm}$. The filtered signal is then focused to produce an image, which is recorded by a liquid-nitrogen cooled CCD. We measure emission intensity $I_1$ for arm 1 open, intensity $I_2$ for arm 2 open, and intensity $I_{12}$ for both arms open, and then calculate
\begin{equation}
 I_{\text{interf}} = (I_{12} - I_1 - I_2) / (2 \sqrt{I_1 I_2}\,)\,.
\label{eqn:I_interf_I}
\end{equation}
In general, for two partially coherent sources located at $\mathbf{r}_1$ and $\mathbf{r}_2$, one has the relation~\cite{Milonni88},
\begin{equation}
 I_{\text{interf}} = \cos \delta \theta(\mathbf{r}_1, \mathbf{r}_2)\, \zeta(\mathbf{r}_1, \mathbf{r}_2)\,,
\label{eqn:I_interf_II}
\end{equation}
where $\delta \theta(\mathbf{r}_1, \mathbf{r}_2)$ is the phase difference of the two sources and $\zeta(\mathbf{r}_1, \mathbf{r}_2)$ is their degree of coherence. In our experimental geometry, there is a small tilt angle $\alpha$ between the image planes of the two arms. As a result, the phase difference
\begin{equation}
\delta \theta(\mathbf{r}_1, \mathbf{r}_2) = q_t y + \phi(\mathbf{r}_1, \mathbf{r}_2)
\label{eqn:delta_theta}
\end{equation}
has a component linear in $y$ --- the coordinate in the direction perpendicular to the tilt axis --- which produces periodic oscillation of $I_{\text{interf}}$. Their period is set by $q_t = 2 \pi \alpha / \lambda$. The coherence function $\zeta(\mathbf{r}_1, \mathbf{r}_2)$ for $\mathbf{r}_1 - \mathbf{r}_2 = \mathbf{\delta x}$ is given by the amplitude of these oscillations,  $A_{\text{interf}}$.

\vskip 5mm \textbf{High-temperature data}

\begin{figure}[tbp]
\centering
\includegraphics[width=8 cm]{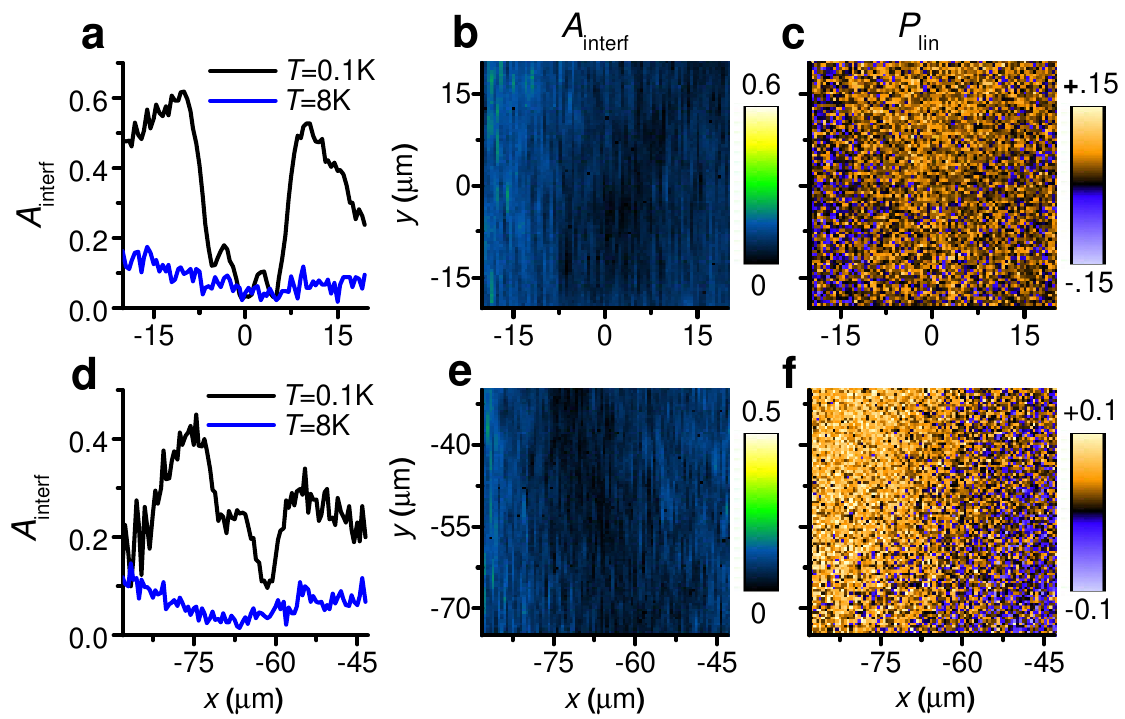}
\caption{High-temperature data: Coherence of indirect excitons in regions of an LBS and the external ring.  (a-c) refer to the region of the LBS in Fig. 2a. (d-f) refer to a region of the external ring in Fig. 2e.
(a,d) $x$ axis cross sections at $y=0$ (a) and $y=-55 \protect\mu$m (d) of $A_{\text{interf}}(x,y)$ at $\protect\delta x = - 2 \protect\mu$m for $T_{\text{bath}} = 0.1\,\text{K}$ (black lines) and 8 K (blue lines). (b,e) $A_{\text{interf}}(x,y)$ at $\protect\delta x = - 2 \protect\mu$m at $T_{\text{bath}} = 8\,\text{K}$. (c,f) Linear polarization of exciton emission at $T_{\text{bath}} = 8\,\text{K}$.}
\label{fig:high-T}
\end{figure}

Figure~\ref{fig:high-T} shows that neither spontaneous coherence nor spin polarization texture is observed at high temperatures in a classical gas in regions of an LBS and the external ring.

\vskip 5mm \textbf{Simulations of interference pattern in the
shift-interferometry experiment}

\begin{figure}[tbp]
\centering
\includegraphics[width=2.0in]{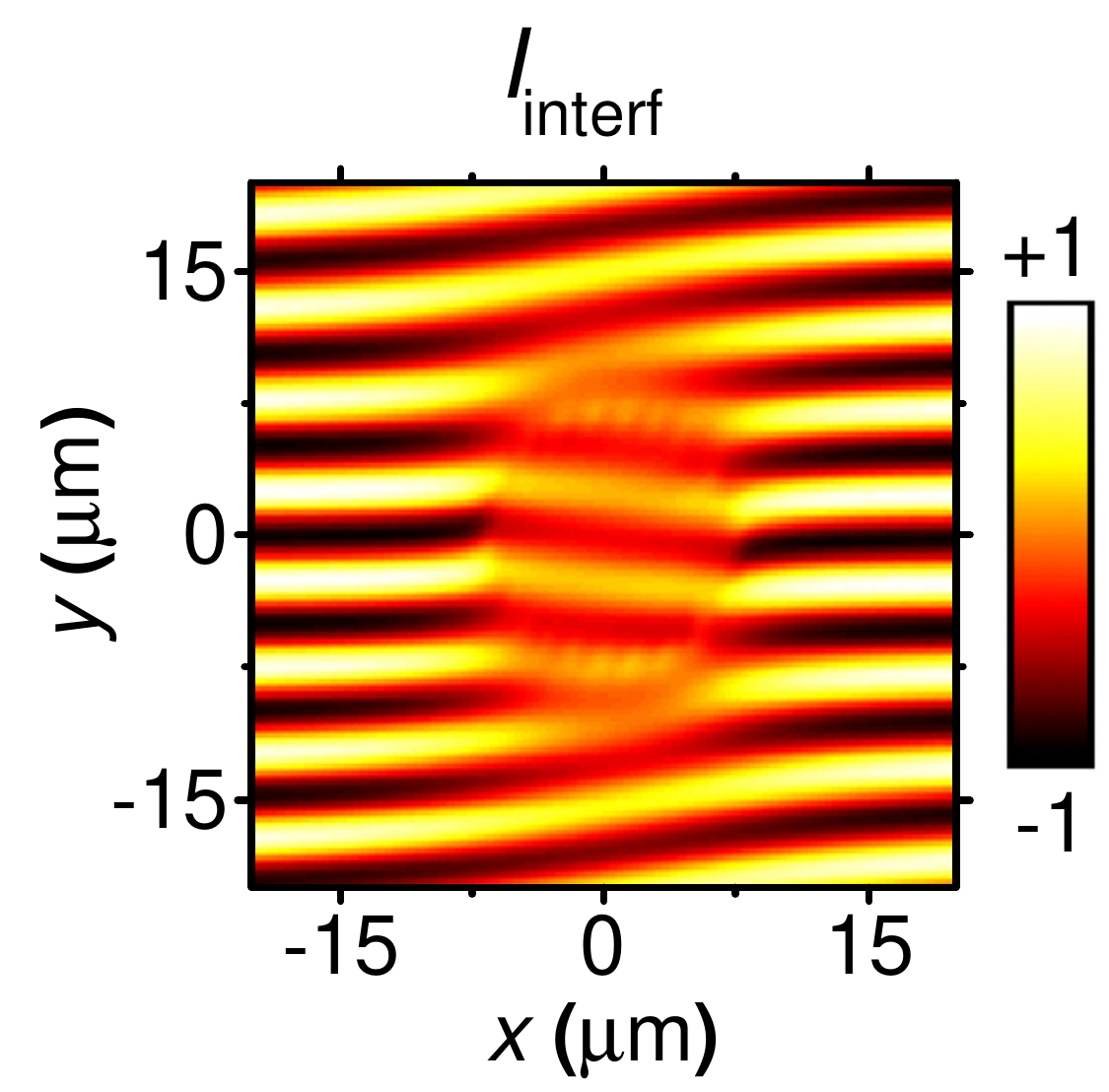}
\caption{Simulations of interference pattern in the shift-interferometry experiments for excitons produced by 12 point sources with radial momenta $q = \pi/2 \,\mu\text{m}^{-1}$ distributed along an 8-$\mu\text{m}$ radius ring.
\label{fig:sim}
}
\end{figure}

The interference pattern in the shift-interferometry experiment is simulated using the formula
\begin{equation}
I_{12} = |\Psi(\mathbf{r}) + e^{i q_t y} \Psi(\mathbf{r} + \mathbf{\delta x})|^2,
\label{eqn:I_sim}
\end{equation}
where the complex function $\Psi(\mathbf{r})$ represents the source amplitude at point $\mathbf{r}$ and $\delta x$ is the in-plane shift.

For a uniform flow of excitons with momentum $\mathbf{q}$, we have $\Psi(\mathbf{r}) = e^{i \mathbf{q r}}$, so that $I_{12} = 2 + 2 \cos (q_t y + \mathbf{q} \mathbf{\delta x})$. In this case, the shift $\delta \phi$ in the phase of interference fringes at $r = r_0$ (Fig. 2b in the main text) gives an estimate of the jump in the exciton momentum $\delta q = \delta \phi / \delta x$.

We also considered a more complicated amplitude function $\Psi$, which was composed of a large number (12) of mutually incoherent radial waves $J_0(q |\mathbf{r} - \mathbf{r}_j|)$ whose centers $\mathbf{r}_j$ were uniformly distributed along a ring. Here $J_0(z)$ is the Bessel function. Each of these sources had the same radial momentum $q$. The simulated interference pattern reproduces the fork-like defects in the measured interference pattern, cf.~Fig.~\ref{fig:sim} and Fig.~5a of the main text. The simulations also show a higher coherence outside the ring region and a higher coherence along the direction of the shift $\mathbf{\delta x}$.

\vskip 5mm \textbf{First-order coherence function}

Coherence of excitons at two separate points $\mathbf{r}_1$ and $\mathbf{r}_2$ of space can be quantified by the density matrix, which is defined by the expectation value
\begin{equation}
\rho_{j k}(\mathbf{r}_1, \mathbf{r}_2) = \langle \psi_j^\dagger(\mathbf{r}_1) \psi_k^{\phantom\dagger}(\mathbf{r}_2) \rangle\,,
\label{eqn:rho_def}
\end{equation}
where $\psi_j^\dagger$ and $\psi_k^{\phantom\dagger}$ are the equal-time creation and annihilation operators of excitons in the four possible $J_z$-states labeled by $j$ and $k$. Coherence of the excitons is imprinted on
their emission:
\begin{equation}
\langle E_j^*(\mathbf{r}_1) E_k^{\phantom{*}}(\mathbf{r}_2) \rangle\
= M_{j k} \rho_{j k}(\mathbf{r}_1, \mathbf{r}_2)\,,
\label{eqn:EE_from_rho}
\end{equation}
where $j$ and $k$ are now restricted to the $J_z = \pm 1$ components and
$E_j$ is the electric field amplitude of the corresponding circular polarization of light. If most of occupied exciton momenta $q$ are much smaller than $2 \pi / \lambda$, we can treat $M_{j k}$ as
constants. Hence, optical measurements can give direct access to determining $\rho_{j k}(\mathbf{r}_1, \mathbf{r}_2)$.

Instead of the left-hand side of Eq.~(\ref{eqn:EE_from_rho}), one traditionally
considers the optical first-order coherence function
\begin{equation}
g_1(\mathbf{r}_1, \mathbf{r}_2) = \sum\limits_{j, k = \pm 1} c_{j k} \langle E_j^*(\mathbf{r}_1) E_k^{\phantom{*}}(\mathbf{r}_2) \rangle\,,
\label{eqn:g_1_def}
\end{equation}
where the coefficients $c_{j k}$ describe the transformation from the circular-polarization basis to the measured polarization state, e.g.,
all $c_{j k} = 1 / 4$ for the $x$-polarization. To elaborate further
on the relation between functions $g_1$ and $\rho_{j k}$,
we make several simplifying assumptions. First, we assume that the system is uniform, so that all observables depend only on the distance $r = |\mathbf{r}_1 - \mathbf{r}_2|$ between the observation points. Second, we neglect spin-orbit coupling and crystal anisotropies, so that we end up with a four-fold degenerate parabolic $\varepsilon_q = \hbar^2 q^2 / (2 m)$ exciton spectrum. Third, we assume that all the four components are equally occupied and completely independent (have no mutual coherence).
%
These assumptions are made primarily because there is no full theoretical understanding of coherence in a gas with spin-orbit coupling, interaction, and spatial inhomogeneity.  From Eqs.~(S6) and (S7) we see that under these assumptions $g_1(r)$ is proportional to the trace $\rho(r)$ of the density matrix.
%

It is also convenient to consider relations between quantities in the momentum space. The Fourier transform of $\rho(r)$ is the momentum distribution $n_{{q}}$. The proportionality between $g_1$ and $\rho$ implies
\begin{equation}
g_1(r) = C_1 \int {d^2 q}\,
e^{i \mathbf{q} \mathbf{r}} n_{{q}}\,,
\label{eqn:g_1_from_n_q}
\end{equation}
where constant $C_1$ provides the normalization $g_1(0) = 1$, which we adopt from now on. Conversely, Eq.~(\ref{eqn:g_1_from_n_q}) indicates that $n_{{q}}$ is proportional to the Fourier transform $\tilde{g}_1(q)$ of $g_1(r)$. This conclusion is very important because function $n_{{q}}$ gives information about interactions, coherence, and phase transitions in the system. Its calculation for interacting particles remains a challenging many-body problem~\cite{Kagan2000qia}. In particular,
no theoretical results for $n_{{q}}$ have been reported for the system under study, where the interaction are dipolar and the spin-orbit coupling is important. For illustrative purposes, we will discuss $n_{{q}}$ and $g_1(r)$ in a \textit{noninteracting} model. In this model $n_{{q}}$ is given by the Bose-Einstein distribution
\begin{equation}
n_{{q}} = \frac{1}{e^{(\varepsilon_{{q}} - \mu) / T} - 1}\,,
\label{eqn:BE}
\end{equation}
where the chemical potential $\mu$ is related to exciton density $n$, temperature $T$, and spectral degeneracy $N = 4$ by
\begin{equation}
\mu = T \ln \left(1 - e^{-n / n_*}\right)\,,
\quad
n_* = \frac{N}{\lambda_{dB}^2}\,.
\label{eqn:mu}
\end{equation}
%
%
%
%
%
%
Here
\begin{equation}
\lambda_{dB} = \sqrt{\frac{2 \pi \hbar^2}{m T}}
\label{eqn:lambda_dB}
\end{equation}
is the thermal de~Broglie wavelength. In turn, the occupation of the $q = 0$ state is given by
\begin{equation}
n_0 = e^{n / n_*} - 1\,.
\label{eqn:n_0}
\end{equation}
As $T$ decreases at fixed $n$, the occupation $n_{{0}}$ of the ground state increases, leading in finite-size systems to condensation of all particles to the ground state.

Let us discuss the consequences of this formula in two limiting cases. First, at high $T$ or low $n$, we deal with a classical gas where
$n_{{q}} \simeq e^{(\mu - \varepsilon_{{q}}) / T} \ll 1$ and Eq.~\eqref{eqn:g_1_from_n_q} yields
\begin{equation}
g_1(r) \simeq e^{-\pi r^2 / \lambda_{dB}^2}.
\quad
\label{eqn:g_1_cl}
\end{equation}
Following the definition of the coherence length $g_1(\xi) = e^{-1}$ in the main text, we conclude that for the classical gas $\xi = \lambda_{dB} / \sqrt{\pi}$. In our system this corresponds to $\xi \approx 0.3\,\mu\text{m}$ at $T = 0.1\,\text{K}$.

In the quantum limit of low $T$ (or high $n$) and $r \gg \lambda_{dB}$, we can replace Eq.~\eqref{eqn:BE} by
\begin{equation}
n_{{q}} \simeq \frac{T}{\varepsilon_{{q}} + |\mu|} \gg 1\,,
\label{eqn:n_q_cl}
\end{equation}
which leads to
\begin{equation}
g_1(r) \propto K_0\left(\frac{r}{\xi_0}\right)\,,
\quad
\xi_0 = \sqrt{\frac{n_0}{4\pi}}\, \lambda_{dB}\,,
\label{eqn:g_1_qu}
\end{equation}
where $K_0(z)$ is the modified Bessel (MacDonald) function. Hence, function $g_1(r)$ decays logarithmically at $\lambda_{dB} \ll r \ll \xi_0$ and then exponentially at larger $r$. The coherence length $\xi$ is somewhat smaller than $\xi_0$ and much larger than $\lambda_{dB}$. As $T$ decreases at fixed $n$, occupation factor $n_0$ and therefore $\xi_0$ grow exponentially with $T$.

The plots of $g_1(r)$ for two representative densities are shown in Fig.~3b by the broken lines. The blue line corresponds to the classical gas where $n_0 \ll 1$ and the black line depicts the quantum gas with $n_0 \approx 5000$. In agreement with the above discussion, crossover from the classical to quantum exciton gas is accompanied by a sharp increase of the coherence length.

\vskip 5mm \textbf{Optical resolution effects}

Experimentally, exciton emission is imaged by an optical system with a finite spatial resolution. Let us denote by $g_1^*(r)$ the coherence function obtained if the resolution is perfect (as assumed in the previous section) and let $g_1(r)$ stand for the observable one. The optical resolution effect can be understood by examining how plane waves with different transverse momenta $q$ pass through the imaging system. The corresponding complex transmission coefficient $f(q)$ is sometimes termed the pupil function. We adopt the following common model~\cite{Hopkins1955tfr}
\begin{equation}
f(q) = \Theta(Q - q) \exp\left(\frac{i}{2}\, D q^2\right)\,,
\quad
Q = k\, \text{NA}\,,
\label{eqn:tilde_f}
\end{equation}
where $k = {2\pi} / {\lambda}$ is the optical wavenumber and $\text{NA}$ is numerical aperture. In our experiments $\text{NA} = 0.27$, which implies $Q = 2.08\,\mu\text{m}^{-1}$. Parameter $D \sim \delta z / k$ accounts for defocusing, with $\delta z$ being the defocusing length. The observed $g_1(r)$ can be expressed as the integral
\begin{equation}
g_1(r) = C_2\, \text{Re}\, \int {d^2 q}\,
e^{i \mathbf{q} \mathbf{r}} f^2(q) \tilde{g}_1^*(q)\,,
\label{eqn:g_1}
\end{equation}
where $C_2$ is another normalization constant. The implications of this model for $g_1(r)$ are as follows.

At high $T$, where the momentum distribution is very broad, $\tilde{g}_1^*(q)$ can be replaced by a constant (see the blue solid line in Fig.~3c of the main text), and so $g_1(r)$ is determined solely by the optical resolution parameters $Q$ and $D$:
\begin{equation}
g_1(r) = C_3 \int\limits_0^Q J_0(q r) \cos D q\, q d q\,.
\label{eqn:PSF}
\end{equation}
The resultant first-order coherence function $g_1(r)$ is characterized by rapidly decaying oscillations, similar to the familiar Airy pattern. In the main text, this structure was referred to as the point-spread function (PSF). A good fit to the experimentally observed high-temperature $g_1(r)$ is obtained for $D = 0.5\,\mu\text{m}^2$, see Fig.~3a of the main text. This corresponds to defocusing length $\delta z \sim D k \approx 4\,\mu\text{m}$, which is reasonable. The width
of the main maximum of the PSF sets the scale of our optical resolution.

At low temperatures and at distances $r \gg 1 / Q$ the optical resolution effects play no role for the shape of the $g_1(r)$ curve. However, the difference in the short-range behavior imposes a different normalization. Hence, we get $g_1(r) = C_4 g_1^*(r)$, where $C_4$ is another constant. The results
are illustrated by the solid lines in Fig.~3b of the main text. In Fig.~3a and 3c we also show that experimental data for $g_1$ and its Fourier transform $\tilde{g}_1$ can be adequately fitted to classical (quantum) gas models at high (low) $T$. However, this fitting should not be overemphasized because of the limited range of the data and
approximations made in the presented theory.


\end{document}